# The Field Perturbation Theory of the Pseudogaps in HTSC

Moshe Dayan

Department of Physics, Ben-Gurion University,

Beer-Sheva 84105, Israel.

# The Field Perturbation Theory of the Pseudogaps in HTSC

Moshe Dayan

Department of Physics, Ben-Gurion University,

Beer-Sheva 84105, Israel.

## Abstract

Here I establish the field perturbation theory of pseudogaps in HTSC. The proposed ground state suggests an internal particle-hole field, which is normal to nesting surfaces, and having twice the Fermi wave-number. It is proved that the system violates momentum conservation by the wave-vector of this internal field. This violation applies to the quasi-particle propagators, as well as to the interactions. Interaction vertices via the Pauli matrix- $\tau_1$ are established. This, in turn, establishes the validity of the pseudogap Hartree self-energy.

1. INTRODUCTION

In a former paper I have developed a field perturbation theory for the pseudogaps in oxide HTSC [1]. The theory is essentially a theory of a condensed phase which is partially analogous to the Nambu-Gorkov theory of superconductivity [2]. The phase transition is a manifestation of a symmetry breaking which is caused by a particle-hole pairing. This particle-hole pairing is a collective coherent phenomenon, as all the pairs carry the same momentum. The symmetry-breaking is triggered by the divergence of electron polarizations at momenta of twice the Fermi momentum, due to nesting states [3]. It has been demonstrated that this latter correlation may take place simultaneously with the Cooper-pair correlation in the superconductive state [4].

Despite some common features, and some analogy of the theories, the superconductive state and the correlation pseudogap state are different in essence. The main difference stems from the correlated species (and also from the momentum of the correlated pair). In the superconductive state, electrons of opposite momentum and spin, pair in a correlated manner- resulting in a gauge symmetry breaking. In the pseudogap state the correlation species are particles and holes, making pairs of charge and spin polarizations. This pairing produces an internal (fixed momentum) field of charge and/or spin polarizations. The fixed momentum is twice the Fermi momentum in size, and is normal to the nesting surfaces. This periodic internal field of the wavenumber $2K_F$ breaks translational symmetry, causing a violation of momentum conservation by $2K_F$. Although this violation could be conjectured from Ref. [1], and although it was briefly discussed in Ref. [4], it was not rigorously established. It is hopefully established in the present paper.

Experiment suggests that, in HTSC, both the superconducting energy gaps and the pseudogaps in the normal state are large. The latter are of the order of 0.2eV, while the former amount usually to several tens of meV. Decades of calculations based on the Nambu-Gorkov theory for various different materials suggest that it is impossible to obtain such values from Fock integrals, given the restrictions imposed on the various relevant parameters [5]. The problem is even more severe when the low electronic density of state in HTSC is taken into account. So far, vast and intensive investigations about the superconductive mechanism in HTSC, have essentially been based on Fock integrals (or their equivalent). This attitude has been proved unsuccessful. I suggest considering Hartree integrals. Indeed, Hartree integrals have been proposed to be the major contributor for the superconductive gaps, and the pseudogaps in HTSC [1, 4]. The reluctance to consider Hartree integrals for the off-diagonal self-energy is understandable, since they are supposed to vanish for the following reason: Hartree diagrams could contribute to the off-diagonal self-energy only by allowing scattering vertices to be spanned by the same Pauli matrices that span the off-diagonal self-energy. Suppose that this matrix is $\tau_1$, then the Hartree Feynman diagrams for the $\tau_1$ self-energy scale with the traces of $\tau_1 m_i$, where $m_i$ is the matrix that define the interaction vertex, and the traces vanish unless $m_i = \tau_1$. In the Nambu-Gorkov theory the interaction vertex scales by $\tau_3$, so that the off-diagonal self-energy should vanish. In the theory of the pseudogap the obvious interaction vertex scales by the unit matrix [1]. However, the nesting sections in the Fermi surface of HTSC, and the conjectured violation of momentum conservation, induce the present author to propose interaction via the $\tau_1$ Pauli matrix in the normal state. In the double correlated state of the combined superconductive-pseudogap problem, two off-diagonal α- Dirac matrices were proposed for interaction vertices- $\alpha_1$ and $\alpha_3$ (in addition to the diagonal $\tau_3$ matrix) [4]. These proposals were made mainly on an intuitive basis. However, in a recent paper, Dayan has shown that an off-diagonal interaction vertex is obtained in any superconducting state, but its related diagram should vanish in all ordinary cases [6]. It has also been conjectured there that, in HTSC, the off-diagonal Hartree diagrams should be significant, due to the involved finite momentum transfer of $2K_F$.

The field in the present paper is based on quasi-particle operators- $\gamma_k$, and $\eta_k$, which define the system excitations, but do not diagonalize the Hamiltonian, when k-parameterization is used over the whole Brillouin zone. This is merely a reflection of the fact that the momentum conservation is violated by $2K_F$. Every momentum **k** by nesting sections of the Fermi surface is paired with momentum $\bar{\mathbf{k}}$, which differs from it by $2K_F$. Consequently, the anti-commutation relations between $\gamma^+_k$ and $\gamma_{\bar{k}}$ (or between $\eta^+_k$ and $\eta_{\bar{k}}$) do not vanish. This does not prevent us from developing field perturbation theory, but suggests some precautions. It is clear that the non-vanishing anti-commutation relations $\{\gamma^+{}_k, \gamma_{\bar{k}}\}$, and $\{\eta^+{}_k, \eta_{\bar{k}}\}$ define the propagators $G_0(\mathbf{k},\bar{\mathbf{k}})$, and $G_0(\bar{\mathbf{k}},\mathbf{k})$, in addition to $G_0(\mathbf{k},\mathbf{k})$, and $G_0(\bar{\mathbf{k}},\bar{\mathbf{k}})$. These propagators, too, are a reflection of the momentum pairing that occurs in the Brillouin zone. It is shown bellow that this pairing leads to off-diagonal interaction vertex, and consequently, to off-diagonal Hartree self-energy.

2. THE GROUND STATE, THE EXCITATIONS, THE FIELD, AND THE HAMILTONIAN.

In Ref. [1] we set the basis for the perturbation field theory of the pseudogaps in the normal state of HTSC, as well as of the correlation gaps in their insulating under-doped counterparts. That analysis was based on an analogy with the Nambu-Gorkov theory of superconductivity. However, there is a serious deviation from this analogy. It is the assumption of off-diagonal interaction vertex, which results in a significant Hartree diagram for the off-diagonal self-energy. The assumption for off-diagonal vertex interaction was based intuitively on the periodic internal fields in the undoped mother materials of HTSC, but was never proved rigorously. In a recent paper we have shown that off-diagonal vertex interactions apply even in ordinary superconductors where momentum is conserved [6]. To establish this, we had to base our field on the Bogoliubov-Valatin excitations which were proved to be in accord with the Wick's theorem. That analysis might suggest that the field operators should be based on the proper excitation of the broken symmetry states, and that transformations to the broken symmetry bases bring with them associated transformations of vertex interactions, where the new vertices conform to the broken symmetry. In the present analysis we also obtain off-diagonal vertices, although here their origin is related to the violation of momentum conservation.

Here, our main goal is to establish the proper field of the under-doped HTSC in their normal state, and to prove the validity of the Hartree diagram for the correlation gaps (and for the pseudogaps in the doped counterparts). Consequently, double correlations [4] are not assumed, and we also disregard disorder, for the sake of simplicity. We shall try to match our functions and notations with Ref. [1], except for some exceptional cases, where we make notice of it. We start by defining the ground state of the condensed phase

$$|\Phi_0> = \prod_{\varepsilon_k<E_F,s} |\Phi_{k,s}> = \prod_{\varepsilon_k<E_F,s} (v_k + u_k c^+_{\bar{k},s} c_{k,s})|0> . \tag{1}$$

In Eq. (1), the states k are within the Fermi surface, $|0>$ is the ground state of the non-condensed phase, and $\bar{\mathbf{k}} = \mathbf{k} - (\mathbf{k}_\perp/|\mathbf{k}_\perp|) \cdot 2\mathbf{k}_F$, where $\mathbf{k}_\perp$ is the component normal to the nesting surface. The parameters $v_k$ and $u_k$ are the probability amplitudes for having the state k full or empty, respectively. Obviously $|v_k|^2 + |u_k|^2 = 1$, and we assume that $v_k$ and $u_k$ are real and positive. Note that the roles of $v_k$ and $u_k$ were reversed relative to Ref. [1], in order to match with the definition of the equivalent parameters in the theory of superconductivity. Particle-hole symmetry considerations imply that $v_{\bar{k}} = u_k$, and $u_{\bar{k}} = v_k$. When the spatial dependence of Eq. (1) is explicitly written, we get

$$|\Phi_{k,s}> = \{v_k + u_k c^+_{\bar{k},s}(x_\perp = 0) c_{k,s}(x_\perp = 0) \exp(i2k_F x_\perp)\}|0>, \tag{2}$$

where $x_\perp$ is the component of **x** in the direction normal to the nesting section by which the state k is located. The second term of Eq. (2) defines charge and spin wave with the wavenumber $2K_F$ in the direction of $k_\perp$. The wave might become CDW if the waves of the two spin states were superimposed with no phase difference, and SDW if they were superimposed anti-phased. Formally, the spin related phase shift could be fixed by the proper choice of the order of the creation operators in the ground state $|0>$. Since the same wave implies to all k's by the discussed section of the Fermi surface, this wave doubles the crystal period in this $k_\perp$ direction. Of course, similar doubling occurs for other nesting sections. Consequently, each part of the Fermi surface which consists of two parallel (nesting) surfaces should produce a violation of momentum conservation by $2K_F$, in the direction normal to the surfaces.

Let us now define the quasi-particle operators

$$\gamma_{k,s} = -u_k c_{k,s} + v_k c_{\bar{k},s} \tag{3a}$$

$$\eta_{k,s} = v_k c^+_{k,s} + u_k c^+_{\bar{k},s} \tag{3b}$$

We immediately see that

$$\gamma_{k,s} | \Phi_0 > = \eta_{k,s} | \Phi_0 > = 0 \tag{4}$$

Eqs. (4) suggest that $\gamma_{k,s}$ and $\eta_{k,s}$ are true annihilation operators of the excitations of the condensed phase. The excitations for k (within the Fermi surface) are defined by

$$\gamma^+_{k,s} | \Phi_{k,s} > = -\gamma^+_{\bar{k},s} | \Phi_{k,s} > = c^+_{\bar{k},s} | 0 > \tag{5a}$$

$$\eta^+_{k,s} | \Phi_{k,s} > = \eta^+_{\bar{k},s} | \Phi_{k,s} > = c_{k,s} | 0 > \tag{5b}$$

The anti-commutation relations between the quasi-particle operators are

$$\{\gamma_{k,s}, \gamma^+_{k',s'}\} = \delta_{s,s'}(\delta_{k,k'} - \delta_{k,\bar{k}'}) \tag{6a}$$

$$\{\eta_{k,s}, \eta^+_{k',s'}\} = \delta_{s,s'}(\delta_{k,k'} + \delta_{k,\bar{k}'}) \tag{6b}$$

$$\{\gamma_{k,s}, \gamma_{k',s'}\} = \{\eta_{k,s}, \eta_{k',s'}\} = \{\gamma^+_{k,s}, \gamma^+_{k',s'}\} = \{\eta^+_{k,s}, \eta^+_{k',s'}\} = 0. \tag{6c}$$

$$\{\gamma_{k,s}, \eta_{k',s'}\} = (\gamma_{k,s}, \eta^+_{k',s'}\} = 0 \tag{6d}$$

Eqs. (4) to (6) guaranty the validity of the Wick's theorem, provided that we consider the propagators $G_0(\mathbf{k}, \bar{\mathbf{k}})$ and $G_0(\bar{\mathbf{k}}, \mathbf{k})$, in addition to $G_0(\mathbf{k}, \mathbf{k})$ and $G_0(\bar{\mathbf{k}}, \bar{\mathbf{k}})$. The second terms of Eqs. (6a) and (6b) result from the pairing of $\mathbf{k}$ and $\bar{\mathbf{k}}$, which effectively divides the reciprocal space into two dependent subspaces.

In accordance with Ref. [1], we define the column vector field operator as

$$\Psi_{k,s} = \begin{pmatrix} -u_k \\ v_k \end{pmatrix} \gamma_{k,s} + \begin{pmatrix} v_k \\ u_k \end{pmatrix} \eta^+_{\bar{k},s}. \tag{7a}$$

Direct calculation shows that the field is identical with the field in Ref. (1)- $\tilde{\Psi}_{k,s}$, namely

$$\Psi_{k,s} = \tilde{\Psi}_{k,s} = \begin{pmatrix} c_{k,s} \\ c_{\bar{k},s} \end{pmatrix} \tag{7b}$$

The Hermitian adjoint of the field is given by

$$\Psi^+_{k,s} = \begin{pmatrix} -u_k & v_k \end{pmatrix} \gamma^+_{k,s} + \begin{pmatrix} v_k & u_k \end{pmatrix} \eta_{\bar{k},s}. \tag{8}$$

One may also define the transposed of each of the fields of Eqs. (7), and (8) by simply transposing the column (row) vector, without changing the $\gamma$ and the $\eta$ operators. The matrix anti-commutation relations are

$$\{\Psi_{k,s}, \Psi^+_{k',s'}\} = \delta_{s,s'}(\delta_{k,k'}I + \delta_{k,\bar{k}'}\tau_1) \tag{9a}$$

$$\{\Psi_{k,s}, \Psi_{k',s'}\} = \{\Psi^+_{k,s}, \Psi^+_{k',s'}\} = 0 \tag{9b}$$

The Hamiltonian density of the non-interacting system is given by

$$H_0'(x) = i\{\Psi^+(x,t) \frac{d}{dt} \Psi(x,t)\}, \tag{10}$$

where

$$\Psi(x,t) = \frac{1}{2} \sum_{k,s} \{M_k \exp(-iE_k t) + N_k \exp(iE_k t)\} \begin{pmatrix} c_{k,s} e^{ikx} \\ c_{\bar{k},s} e^{i\bar{k}x} \end{pmatrix}. \tag{11}$$

In Eq. (11), $M_k = \begin{bmatrix} u_k^2 & -u_k v_k \\ -u_k v_k & v_k^2 \end{bmatrix}$, and $N_k = \begin{bmatrix} v_k^2 & u_k v_k \\ u_k v_k & u_k^2 \end{bmatrix}$. The time dependence of $\Psi(x,t)$ in Eq. (11) is obtained from the time dependence of the operators $\gamma_{k,s}$ in the interaction picture, which is given by $\exp(-iE_k t)$, and the time dependence of $\eta_{k,s}^+$, which is given by $\exp(iE_k t)$. Thus,

$$H_0' = \frac{1}{2} \sum_{k,k',s,s'} E_k \{\delta_{k,k'} \delta_{s,s'} \left( c_{k',s'}^+ e^{-ik'x}, \ c_{\bar{k}',s'}^+ e^{-i\bar{k}'x} \right) [M_k^T M_k - N_{k'}^T N_k] \begin{pmatrix} c_{k,s} e^{ikx} \\ c_{\bar{k},s} e^{i\bar{k}x} \end{pmatrix} \}$$

$$+ \frac{1}{2} \sum_{k,k',s,s'} E_k \{\delta_{\bar{k},k'} \delta_{s,s'} \left( c_{k',s'}^+ e^{-ik'x}, \ c_{\bar{k}',s'}^+ e^{-i\bar{k}'x} \right) [M_{k'}^T M_k - N_{k'}^T N_k] \begin{pmatrix} c_{k,s} e^{ikx} \\ c_{\bar{k},s} e^{i\bar{k}x} \end{pmatrix} \}. \tag{12}$$

One can easily verify that $M_k^T M_k = M_k$, $N_k^T N_k = N_k$, $M_{\bar{k}}^T M_k = -(2u_k v_k)\tau_1 M_k$, and $N_{\bar{k}}^T N_k = (2u_k v_k)\tau_1 N_k$. Consequently,

$$H_0' = \frac{1}{2} \sum_{k,s} E_k \{ \left( c_{k,s}^+ e^{-ikx}, \ c_{\bar{k},s}^+ e^{-i\bar{k}x} \right) [M_k - N_k] \begin{pmatrix} c_{k,s} e^{ikx} \\ c_{\bar{k},s} e^{i\bar{k}x} \end{pmatrix}$$

$$- \frac{1}{2} \sum_{k,s} E_k \{ 2u_k v_k \left( c_{\bar{k},s}^+ e^{-i\bar{k}x}, \ c_{k,s}^+ e^{-ikx} \right) \tau_1 [M_k + N_k] \begin{pmatrix} c_{k,s} e^{ikx} \\ c_{\bar{k},s} e^{i\bar{k}x} \end{pmatrix} \tag{13}$$

Let us denote $\varepsilon_k = E_k (u_k^2 - v_k^2)$, and $\Lambda_k = -E_k 2 u_k v_k$, and obtain

$$H_0'(x) = \frac{1}{2} \sum_{k,s} \{\varepsilon_k \tilde{\Psi}_{k,s}^+ \tau_3 \tilde{\Psi}_{k,s} + \Lambda_k [\tilde{\Psi}_{k,s}^+ \tilde{\Psi}_{k,s} + \tilde{\Psi}_{k,s}^+(x) \tau_1 \tilde{\Psi}_{k,s}(x)]\} \tag{14}$$

The form of Eq. (14) suggests immediately that, when k is out of the Fermi surface, then $\varepsilon_k$ is indeed the energy of the excitations $c^+_{k,s}|0>$, and $c^-_{\bar{k},s}|0>$. From the definitions of $\varepsilon_k$, one gets the equivalent of the BCS equations $u_k^2 = \frac{1}{2}\{1+(\varepsilon_k/E_k)\}$, and $v_k^2 = \frac{1}{2}\{1-(\varepsilon_k/E_k)\}$. From the definition of $\Lambda_k$, one gets immediately the relation $E_k^2 = (\varepsilon_k^2 + \Lambda_k^2)$. The second term in Eq. (14) is merely a shift of the Fermi level. This shift might be offset by a shift of the frequency scale, a shift which is discussed in the last section of this paper. The last term in Eq. (14) is the Hartree-Fock term of the condensation. The spatial dependence is denoted only in this term because it vanishes in the others. Explicitly, the condensation term is

$$\frac{1}{2}\sum_{k,s}\Lambda_k \tilde{\Psi}^+_{k,s}(x)\tau_1\tilde{\Psi}_{k,s}(x) = \frac{1}{2}\sum_{k,s}\Lambda_k\{c^+_{k,s}c_{\bar{k},s}\exp(-i2k_Fx_\perp) + c^+_{\bar{k},s}c_{k,s}\exp(i2k_Fx_\perp)\}. \quad (15)$$

A direct application of $H_0'$ on the ground state yields

$$H_0'|\Phi_0> = \{\sum_{k<k_F}-(E_k - \Lambda_k)\}|\Phi_0>. \quad (16)$$

We wish to set the zero energy level so that the eigenvalue of the ground state is zero, Therefore we add a constant and write

$$H_0(x) = \frac{1}{2}\sum_{k,s}\{\varepsilon_k\tilde{\Psi}^+_{k,s}\tau_3\tilde{\Psi}_{k,s} + \Lambda_k[\tilde{\Psi}^+_{k,s}\tilde{\Psi}_{k,s} + \tilde{\Psi}^+_{k,s}(x)\tau_1\tilde{\Psi}_{k,s}(x)] + (E_k - \Lambda_k)\}. \quad (17)$$

Applying $H_0$ on the excitations $\gamma^+_{k,s}|\Phi_0>$ and $\eta^+_{k,s}|\Phi_0>$ (when k is inside the Fermi surface) yields

$$H_0\gamma^+_{k,s}|\Phi_0> = (E_k + \Lambda_k)\gamma^+_{k,s}|\Phi_0> \quad (18)$$

$$H_0\eta^+_{k,s}|\Phi_0> = (E_k - \Lambda_k)\eta^+_{k,s}|\Phi_0> \quad (19)$$

Eqs. (18) and (19) prove that $\gamma^+_{k,s}|\Phi_0>$ and $\eta^+_{k,s}|\Phi_0>$ are indeed the excitations of the condensed system. When the shift of the Fermi energy is readjusted as discussed, one finds that both their eigenvalues become $E_k$. This is so because Eqs. (5) suggest that, while $\gamma^+_{k,s}|\Phi_0>$ may be considered as a quasi-particle, $\eta^+_{k,s}|\Phi_0>$ should be considered as a quasi-hole.

3. THE PROPAGATOR, THE FEYNMAN-DYSON'S PERTURBATION EXPANSION, AND THE OFF-DIAGONAL VERTEX INTERACTION.

The definitions and derivations of the former section set up the foundations for deriving field perturbation theory for the discussed system. Let us start with the propagator $G_0$,

$$G_0(k,t) = -i <\Phi_0 | T\{\Psi_{k,s}(t), \Psi^+_{k,s}(0)\} | \Phi_0> , \qquad (20)$$

where $T$ is the time ordering operator. Using Eqs. (7) and (8) we get

$$G_0(k,t) = -i <\Phi_0 | M_k \exp(-iE_k t)\Theta(t) - N_k \exp(iE_k t)\Theta(-t) | \Phi_0>. \qquad (21)$$

The time Furrier transform is carried out after representing the step functions $\Theta$ by their integral forms. We get

$$G_0(k,\omega) = \frac{M_k}{\omega - E_k + i\delta} + \frac{N_k}{\omega + E_k - i\delta}, \qquad (22)$$

which, after substituting $M_k$ and $N_k$, becomes

$$G_0(k,\omega) = (\omega^2 - E_k^2 + i\delta)^{-1}\{\omega I + \varepsilon_k \tau_3 + \Lambda_k \tau_1\}. \qquad (23)$$

As expected, Eq. (23) has the analogous form as its counterpart in the theory of superconductivity. However, the analogy is broken by the existence of Eq. (9b), which suggests the existence of $G_0(\bar{k},k,\omega)$, which does not have an analog in the theory of superconductivity. The propagator $G_0(\bar{k},k,\omega)$ is given by

$$G_0(\bar{k},k,t) = -i <\Phi_0 | T\{\Psi_{\bar{k},s}^-(t), \Psi_{k,s}^+(0)\} | \Phi_0>$$

$$= -i <\Phi_0 | \tau_1 M_k \exp(-iE_k t)\Theta(t) - \tau_1 N_k \exp(iE_k t)\Theta(-t) | \Phi_0>. \quad (24)$$

This immediately suggests

$$G_0(\bar{k},k,\omega) = \tau_1 G_0(k,\omega), \quad (25a)$$

and similarly

$$G_0(k,\bar{k},\omega) = G_0(k,\omega)\tau_1, \quad (25b)$$

$$G_0(\bar{k},\bar{k},\omega) = G_0(\bar{k},\omega) = \tau_1 G_0(k,\omega)\tau_1. \quad (25c)$$

All the necessary physical information we need exists in $G(k,\omega)$, the propagator of the interacting system. This propagator is obtained by field perturbation theory, employing the interaction Hamiltonian, which is given by

$$H_i = \frac{1}{8}\sum_{k,k',s,s',q} V_q \tilde{\Psi}_{k'-q,s'}^+ \tilde{\Psi}_{k',s'} \tilde{\Psi}_{k+q,s}^+ \tilde{\Psi}_{k,s} - \frac{1}{2}\sum_{k,s} \Lambda_k [\tilde{\Psi}_{k,s}^+ \tilde{\Psi}_{k,s} + \tilde{\Psi}_{k,s}^+(x)\tau_1 \tilde{\Psi}_{k,s}(x)]. \quad (26)$$

The subtraction of the second term, which is the Hartree-Fock term, compensates for the addition of the same term in $H_0$. Thus, when the perturbation is performed within the above Hartree-Fock approximation it should yield zero, because it is built in the unperturbed system. When the perturbation is done according to the Wick's theorem, and Eqs. (4) and (9), the expansion of $G(k,\omega)$ includes elements like $G_0(k,\omega)$, $G_0(k,\bar{k},\omega)$, $G_0(\bar{k},k,\omega)$, $G_0(\bar{k},\bar{k},\omega)$, and the interactions between them. Eqs. (25) Suggest that one can eliminate $G_0(k,\bar{k},\omega)$, $G_0(\bar{k},k,\omega)$ and $G_0(\bar{k},\bar{k},\omega)$ from the expansion, by simply replacing them with $G_0(k,\omega)$, provided that a $\tau_1$ matrix is substituted at every vertex where $\bar{k}$ is replaced by $k$.

The perturbation expansion of $G(k,t)$ is a sum of terms that are "dressed" by self-energy diagrams of all orders. Here we wish to demonstrate such contributions of the first and of the second order in the interaction Hamiltonian. They are given by

$$G_1(k,s,t) = \frac{i\int dt_1 <\Phi_0 | T\{H_i'(t_1)\Psi_{k,s}(t)\Psi_{k,s}^+(0)\} | \Phi_0 >}{2 <\Phi_0 | U | \Phi_0 >} , \quad (27a)$$

$$G_2(k,s,t) = \frac{i\int dt_1 dt_2 <\Phi_0 | T\{H_i'(t_1)H_i'(t_2)\Psi_{k,s}(t)\Psi_{k,s}^+(0)\} | \Phi_0 >}{2 <\Phi_0 | U | \Phi_0 >} , \quad (27b)$$

Where $U$ is the time development operator, and

$$H_i'(t_1) = \frac{1}{8} \sum_{k_1,k_1',s_1,s_1',q_1} V_{q_1}(t_1) \Psi_{k_1'-q_1,s_1'}^+(t_1) \Psi_{k_1',s_1'}(t_1) \Psi_{k_1+q_1,s_1}^+(t_1) \Psi_{k_1,s_1}(t_1) , \quad (28a)$$

$$H_i'(t_2) = \frac{1}{8} \sum_{k_2,k_2',s_2,s_2',q_2} V_{q_2}(t_2) \Psi_{k_2'-q_2,s_2'}^+(t_2) \Psi_{k_2',s_2'}(t_2) \Psi_{k_2+q_2,s_2}^+(t_2) \Psi_{k_2,s_2}(t_2) . \quad (28b)$$

The self-energy of $G_1(k,t)$ contains only bare interactions, whereas the interactions in $G_2(k,t)$ are screened. Screened interactions are obtained by two contractions between the fields of one vertex of $H'_i(t_1)$ and the fields of one vertex of $H'_i(t_2)$. A momentum defined interaction is an interaction with the same momentum parameters at its two vertices. While bare interactions are always momentum defined, screened interactions may or may not be momentum defined. A momentum undefined interaction carries $\mathbf{q}$ momentum at one vertex, and $\bar{\mathbf{q}}$ momentum at its other vertex, where $|\mathbf{q} - \bar{\mathbf{q}}| = 2k_F$. Let us demonstrate this in second order interaction, using the notations of Eqs. (28). We contract $\Psi^+_{k_2+q_2,s_2}(t_2)$ with $\Psi_{k'_1,s'_1}(t_1) = \Psi^-_{(\bar{k}_2+q_2),s'_1}(t_1)$, and $\Psi^+_{k'_1-q_1,s'_1}(t_1)$ with $\Psi_{k_2,s_2}(t_2)$. This implies that $\mathbf{q}_1 = \mathbf{q}_2 - 2k_F \mathbf{k}_{2\perp}/|k_{2\perp}| = \bar{\mathbf{q}}_2$, $s_2 = s'_1$, and the contractions are equal to $\tau_1 G_0(k_2+q,t_1,t_2) \times G_0(k_2,t_2,t_1)$. This product of propagators (when properly integrated) is known as the simplest electron polarization. The associated interaction diagram is shown in Fig. 1a. Obviously, the interaction violates momentum conservation by $2k_F$, due to the single $\tau_1$ vertex of its polarization. This is actually a rule: every $\tau_1$ vertex indicates that the vertex violates momentum conservation by $2k_F$. Consequently, when the number of the $\tau_1$ matrices within the interior of an interaction line (not including the end vertices) is even, the interaction is momentum defined. When that number is odd, the interaction is not momentum defined- it is off-diagonal interaction. From Eqs.(28) we conclude that originally the interactions are terminated by $I$ vertices. However, any $I$ vertex which causes momentum transfer $\bar{\mathbf{q}}$ between the scattered electrons, may be converted into a $\tau_1$ vertex, with momentum transfer $\mathbf{q}$ between the scattered electrons. Thus, the discussed momentum undefined interaction could be transformed into a *formally momentum defined* (FMD) interaction by converting one of its $I$ vertices into a $\tau_1$ vertex. This transformation into formally momentum defined interaction does not change the essential deviation from momentum conservation, but merely shifts it into the $\tau_1$ vertex. This is shown in Fig. 1b. Notice that both the originally momentum defined interactions and the transformed FMD interactions have an overall even number of $\tau_1$ vertices (including edge vertices).

The above discussion suggests that we could keep the convenient convention of the usual Feynman rules, provided that we allow $\tau_1$ vertices, and keep the following rules:
1) The interactions, as well as the quasi-particle propagators, are 2x2 matrices. An interaction could also be off-diagonal (namely- proportional to $\tau_1$).
2) For Feynman diagrams, use only **momentum defined propagators and interactions**. When an interaction is off-diagonal, it becomes formally momentum defined when it is terminated by one $I$ vertex and one $\tau_1$ vertex. In this form it can be used in a Feynman diagram. Thus, every interaction line in a Feynman diagram includes an even number of $\tau_1$ matrices (including the edge vertices).

3) Assume momentum conservation at every $I$ vertex, and a deviation by $|\mathbf{q} - \bar{\mathbf{q}}| = 2k_F$ at every $\tau_1$ vertex.

Thus, we may have momentum defined interactions that are terminated by two $I$ vertices, and momentum defined interactions that are terminated by two $\tau_1$ vertices. Assuming that the latter were produced from vertex conversion of $\bar{q}$- defined interactions of the former type, they are given by $\tilde{V}_{\tau\tau,q} = \tau_1(1 - V_{\bar{q}}\tau_1\tilde{\Pi}_q\tau_1)V_{\bar{q}}\tau_1$, while the former are given by $\tilde{V}_{II,q} = V_q(1 - \tilde{\Pi}_q V_q)$, where $\tilde{\Pi}_q$ is the sum of all polarizations. The momentum undefined interactions are given by $\tilde{V}_{I\tau,q\bar{q}} = -V_q\tilde{\Pi}_q\tau_1 V_{\bar{q}}$, which should be converted into formally momentum defined ($\tilde{V}_{I\tau,q} = -V_q\tilde{\Pi}_q\tau_1 V_{\bar{q}}\tau_1$) by multiplying its right hand side with $\tau_1$. The Dyson equation for $\tilde{\Pi}_q$ is

$$\tilde{\Pi}_q = \Pi_q + \tau_1\Pi_{\bar{q}}\tau_1 - \Pi_q(V_q + \tau_1 V_{\bar{q}}\tau_1)\tilde{\Pi}_q - \tau_1\Pi_{\bar{q}}\tau_1(V_q + \tau_1 V_{\bar{q}}\tau_1)\tilde{\Pi}_q \tag{29a}$$

$$\tilde{\Pi}_q = (\Pi_q + \tau_1\Pi_{\bar{q}}\tau_1)[1 + (\Pi_q + \tau_1\Pi_{\bar{q}}\tau_1)(V_q + \tau_1 V_{\bar{q}}\tau_1)]^{-1}, \tag{29b}$$

where $\Pi_q$ is the irreducible polarization. The interaction relevant to the off-diagonal Hartree diagram is

$$V_H = \tilde{V}_{\tau\tau,q} + \tilde{V}_{I\tau,q} = \tau_1 V_{\bar{q}}\tau_1[1 - \tilde{\Pi}_q(\tau_1 V_{\bar{q}}\tau_1 + V_q)]$$

$$= \tau_1 V_{\bar{q}}\tau_1[1 + (V_q + \tau_1 V_{\bar{q}}\tau_1)(\Pi_q + \tau_1\Pi_{\bar{q}}\tau_1)]^{-1} \tag{30}$$

In the small q limit, this interaction vanishes as $V_H \xrightarrow{q \to 0} \dfrac{q^2}{\bar{q}^2(\Pi_q + \tau_1\Pi_{\bar{q}}\tau_1)}$. Thus, off-diagonal Hartree diagrams vanish if only screened Coulomb interaction is taken into account. One has to consider el-phonon-el interaction, too.

For the inclusion of the el-phonon-el interaction, one must define the phonon field and its interactions with electrons. This certainly would be an essential and comprehensive analysis. However, here we make a short-cut by adding the elemental el-phonon-el bare interaction- $\sum_{\lambda}\left|g_{q,\delta k,\lambda}\right|^{2} D_{q,\lambda}$, to every $V_q$ in the interaction Hamiltonians of Eqs. (28). Here $\lambda$ is the phonon mode, $D_{q,\lambda}$ is the propagator of the bare phonon, and $g_{q,\delta k,\lambda}$ is the bare matrix element for its interaction with electrons that scatter with momentum transfer of $\delta \mathbf{k}$. This momentum transfer is given by $\delta \mathbf{k} = \mathbf{Q}_m + \mathbf{q}$, where $\mathbf{Q_m}$ is a vector of the reciprocal lattice. One should notice that this interaction is time dependent, whereas the bare Coulomb interaction is immediate in the non-relativistic limit. We also stress that we assume a semi-metallic model, where the positive charge of the basis of ions in the unit cell is neutralized by the negative charge of the electrons. In such a model the bare longitudinal zero-q acoustic phonon has a finite frequency, which is corrected to become zero only by electron screening. Although this model might look inappropriate for under-doped HTSC, it should be noticed that we assume it only for the bare phonons, before perturbation and renormalization.

The first question that should be addressed when attempting this treatment is whether to assume the regular lattice periodicity, or the "anti-ferromagnetic" periodicity even for the lattice. The latter periodicity is described by transforming the vectors of the basal plane, from $(1,0,0)x(0,1,0)$, into $(1,1,0)x(1,\bar{1},0)$, causing its area to double. The reciprocal lattice is in accordance with the nesting Fermi surface, with a period of $2K_F$. Our treatment so far suggests that this is indeed the periodicity of the electronic system, but for the el-phonon interaction the lattice periodicity is the relevant one. Experiments show that the lattice periodicity of the under-doped orthorhombic $La_{2-x}(Ba,Sr)_x CuO_4$ is indeed in accord with the "anti-ferromagnetic" period (in addition to its transformation from tetragonal to orthorhombic) [7-9]. However, $YBa_2Cu_3O_{7-x}$ has the regular periodicity (although in the orthorhombic symmetry) [10,11]. Bismuth, Thallium, and Lead Cupper oxides may have both structures [12]. A preliminary study of the two versions of the lattice periodicity suggests that their off-diagonal Hartree potential- $V_H$, might be different. Both versions deserve study. However, in the preliminary treatment of the present paper, we assume only the regular periodicity.

Let us denote

$$V_q^t = V_q + \sum_\lambda |g_{q,\delta k,\lambda}|^2 D_{q,\lambda} \tag{31a}$$

$$\tau_1 V_{\bar{q}}^t \tau_1 = \tau_1 V_{\bar{q}} \tau_1 + \tau_1 \sum_\lambda |g_{\bar{q},\delta k,\lambda}|^2 D_{\bar{q},\lambda} \tau_1. \tag{31b}$$

The bare el-phonon matrix element is given by [13,14]

$$g_{q,\delta k,\lambda} = -i \sum_\alpha \sqrt{\frac{N}{2M_\alpha \Omega_{q,\lambda}}} \delta \mathbf{k} \cdot \mathbf{e}_{\alpha,\lambda}(\mathbf{q}) V_\alpha(\delta \mathbf{k}), \tag{32}$$

where $N$ is the number of unit cells, $M_\alpha$ is the mass of the $\alpha$-th ion in a unit cell, $\mathbf{e}_{\alpha,\lambda}$ is a unit vector in the direction of the polarization of the $\alpha$-th ion in the $\lambda$-th mode, and $\Omega_{q,\lambda}$ is the bare phonon frequency. The Fourier transform of the potential of the $\alpha$-th ion is given by

$$V_\alpha(\delta \mathbf{k}) = \int d^3 r V_\alpha(\mathbf{r} - \mathbf{R}_\alpha) \exp(-i \delta \mathbf{k} \cdot \mathbf{r})], \tag{33}$$

where $\mathbf{R}_\alpha$ is the position of the $\alpha$-th ion (relative to a chosen origin in the cell). When the Hartree diagram is considered, the relevant interaction is $V_H^t = \tilde{V}_{\tau\tau,q}^t + \tilde{V}_{I\tau\tau,q}^t$, with $q = 0$, and $\omega = 0$. The off-diagonal Hartree diagram is depicted in Fig. 1a of Ref. [1]. Its associated interaction is given by,

$$V_H^t(q = \omega = 0) = U_H = (\tau_1 V_{\bar{q}}^t \tau_1)[1 + (V_q^t + \tau_1 V_{\bar{q}}^t \tau_1)(\Pi_q + \tau_1 \Pi_{\bar{q}} \tau_1)]^{-1}. \tag{34}$$

A fast evaluation of $U_H$ suggests that it is finite and negative, provided that $|(V_q^t + \tau_1 V_{\bar{q}}^t \tau_1)(\Pi_q + \tau_1 \Pi_{\bar{q}} \tau_1)| < 1$.

The interaction for the Fock integral is not limited to $q = \omega = 0$, but is momentum and energy integrated. It is given by

$$V_F^t(\omega,q) = U_F(\omega,q) = \tilde{V}_{II,q}^t + \tilde{V}_{\tau\tau,q}^t + \tilde{V}_{I\tau,q}^t = \frac{V_q^t + \tau_1 V_{\bar{q}}^t \tau_1}{1 + (V_q^t + \tau_1 V_{\bar{q}}^t \tau_1)(\Pi_q + \tau_1 \Pi_{\bar{q}} \tau_1)} \quad . \quad (35)$$

## 4. CONCLUSIONS.

The above analysis seems to put the treatment of Ref. [1] on firmer grounds, as a perturbation field theory. It proves the validity of the field, the validity of the off-diagonal vertex interaction, and consequently, of the Hartree order parameter- $\Lambda_H$. During the course of the analysis, we have established the existence of an internal field, which causes a violation of momentum conservation by $2k_F$. We showed that this violation applies to electron propagators, as well as to their interactions. The analysis also provides a better insight into the interaction, and a better evaluation of its strength.

According to Eq. (18) of Ref. [1]

$$\Lambda = 2E_m \exp(\frac{1-f}{2U_H N_0}) \quad (36)$$

where $2E_m$ is the energy range in which nesting occurs, $N_0$ is the density of states at the Fermi level, in the un-condensed phase, and $f = \Lambda_F / \Lambda_H$, is the ratio between the off-diagonal Fock and Hartree integrals. We assumed that $f \ll 1$. We have shown that the argument of the exponent is negative, and we assume that its absolute value is not much larger than unity, resulting in a large order parameter. Eq. (34) establishes the finiteness and the proper sign of $U_H$, which is a key parameter in Eq. (36).

The Fock integral too survives our re-analysis. Eq. (23) in Ref. [1] shows that the integrand is proportional to $\text{Re}\{\Lambda/\sqrt{v^2 - \Lambda^2}\}$, where $v$ is the frequency parameter of integration. This suggests that the integrand become significant only at energies larger than $\Lambda$, which in turn is larger than the Debye energy. Thus, contrary to the Hartree integral, the Fock integral (at energies lower than $\Lambda$) is determined mainly by the Coulomb interaction. In Eq. (24) of Ref. [1] it was evaluated as

$$\Lambda_F = \frac{1}{2} N_0 \overline{U}_F \Lambda \ln \frac{E_m}{|\Lambda|} \tag{37}$$

where $\overline{U}_F$ is a momentum average of the interaction $U_F$.

The Hamiltonian of Eq. (17) includes a term that shifts the energy scale, and could be offset by a shift of the Fermi level. Should this shift be done arbitrarily, or might it already been built into the formalism? Indeed, our formalism allows a shift of the frequency scale via the diagonal Hartree integral. This is originated by $G_0(k,\overline{k},\omega)$, whose diagonal self-energy equals the off-diagonal self-energy of $G_0(k,\omega)$. This is the reason for having the $I$ and the $\tau_1$ components of the unperturbed Hamiltonian of Eq. (14) pre-factored by the same parameter- $\Lambda_k$.

The present analysis is based on the field $\widetilde{\Psi}_{k,s}$ which is defined in Eqs. (7), and is equivalent to the Nambu field for superconductivity. Evidently, there is an essential difference between the two problems. In the present problem one gets off-diagonal interaction vertex even with the Nambu field $\widetilde{\Psi}_{k,s}$, where in the superconductive problem one has to define a different field (the coherence field) in order to get off-diagonal vertex [6]. It is interesting to show that coherence fields may be defined here, too. Let us define

$$\hat{\Psi}_{k,s} = \begin{pmatrix} -u_k \\ v_k \end{pmatrix} \gamma_{k,s} - \begin{pmatrix} v_k \\ u_k \end{pmatrix} \eta^+_{k,s} = E_k^{-1}(\varepsilon_k \tau_3 + \Lambda_k \tau_1) \widetilde{\Psi}_{k,s}. \tag{38}$$

The interaction vertex, which is $I(k',k) = V_{k',k} \widetilde{\Psi}^+_{k',s'} I \widetilde{\Psi}_{k,s}$, is written in terms of the coherence field as

$$I(k',k) = V_{k',k} \hat{\Psi}^+_{k',s'} \{\delta_{k',k} I + \delta_{\overline{k}',k}[(-1+8u_k^2 v_k^2)\tau_1 - 4u_k v_k(u_k^2 - v_k^2)\tau_3]\} \hat{\Psi}_{k,s}$$

$$= V_{k',k} \hat{\Psi}^+_{k',s'} \{\delta_{k',k} I + \delta_{\overline{k}',k} E_k^{-2}[-(\varepsilon_k^2 - \Lambda_k^2)\tau_1 + 2\Lambda_k \varepsilon_k \tau_3]\} \hat{\Psi}_{k,s}. \tag{39}$$

Thus, the part which does not conserve momentum acquired a $\tau_3$ term, in addition to the $\tau_1$ term.

To conclude we should notice that we have assumed strong under-doping in order to treat somewhat ideal Fermi surfaces where nesting states dominate. We have also disregarded disorder for the sake of simplicity. The present analysis also has not considered superconductivity, and the related double correlations [4]. When these practical aspects of HTSC are considered, one should add the relevant modifications as discussed in Refs. [1] and [4].

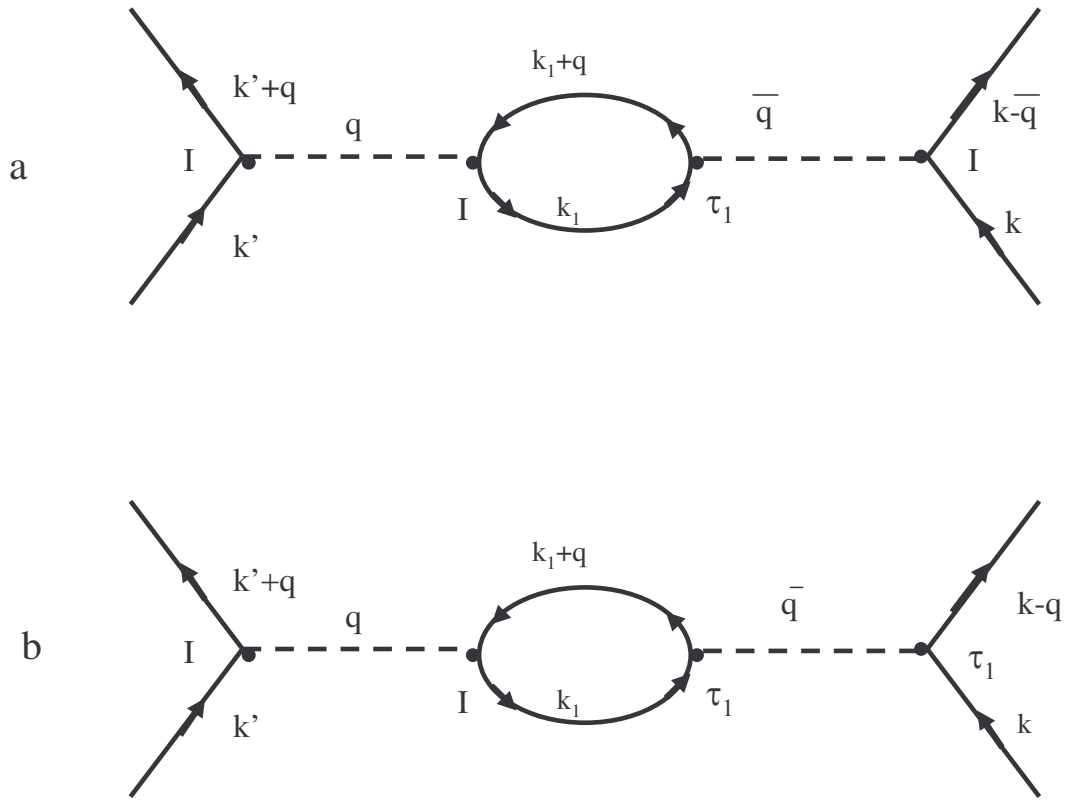

Fig. 1

a. An example of screened interaction which violates momentum conservation by $2K_F$. Notice the $\tau_1$ vertex at which the violation of momentum conservation is originated.

b. The same interaction is converted to **formally** momentum conserving by replacing an *I* vertex with a $\tau_1$ vertex. Notice that the difference between the two diagrams is only formal (but not essential).

.